# Physics-informed motion registration of lung parenchyma across static CT images

Sunder Neelakantan
*Department of Biomedical Engineering*
*Texas A&M University*
College Station, TX, USA
Email: sundern@tamu.edu

Tanmay Mukherjee
*Department of Biomedical Engineering*
*Texas A&M University*
College Station, TX, USA
Email: tanmaymu@tamu.edu

Kyle J. Myers
*Hagler Institute for Advanced Study*
*Texas A&M University*
College Station, TX, USA
Email: drkylejmyers@gmail.com

Rahim Rizi
*Department of Radiology*
*Perelman School of Medicine*
*University of Pennsylvania*
Philadelphia, PA, USA
Email: rizi@pennmedicine.upenn.edu

Reza Avazmohammadi
*Department of Biomedical Engineering*
*Texas A&M University*
College Station, TX, USA
Email: rezaavaz@tamu.edu

*Abstract*-Lung injuries, such as ventilator-induced lung injury and radiation-induced lung injury, can lead to heterogeneous alterations in the biomechanical behavior of the lungs. While imaging methods, e.g., X-ray and static computed tomography (CT), can point to regional alterations in lung structure between healthy and diseased tissue, they fall short of delineating timewise kinematic variations between the former and the latter. Image registration has gained recent interest as a tool to estimate the displacement experienced by the lungs during respiration via regional deformation metrics such as volumetric expansion and distortion. However, successful image registration commonly relies on a temporal series of image stacks with small displacements in the lungs across succeeding image stacks, which remains limited in static imaging. In this study, we have presented a finite element (FE) method to estimate strains from static images acquired at the end-expiration (EE) and end-inspiration (EI) timepoints, i.e., images with a large deformation between the two distant timepoints. Physiologically realistic loads were applied to the geometry obtained at EE to deform this geometry to match the geometry obtained at EI. The results indicated that the simulation could minimize the error between the two geometries. Using four-dimensional (4D) dynamic CT in a rat, the strain at an isolated transverse plane estimated by our method showed sufficient agreement with that estimated through non-rigid image registration that used all the timepoints. Through the proposed method, we can estimate the lung deformation at any timepoint between EE and EI. The proposed method offers a tool to estimate timewise regional deformation in the lungs using only static images acquired at EE and EI.

## I. INTRODUCTION

Lung injuries, such as inflammation [1], ventilator-induced lung injury (VILI) [2]-[4], and radiation-induced lung injury (RILI) [5], [6],can lead to heterogeneous alterations in the biomechanical behavior of the lungs. Imaging methods, such as X-ray imaging and static computed tomography (CT) scans, are available for the clinical assessment of lung injuries. While such methods provide morphological information about the lung structure, they cannot directly provide information about the kinematics of lung parenchyma. Image registration [7]-[12] has gained recent interest as a post-processing tool to quantify the kinematic behavior of lung parenchyma.

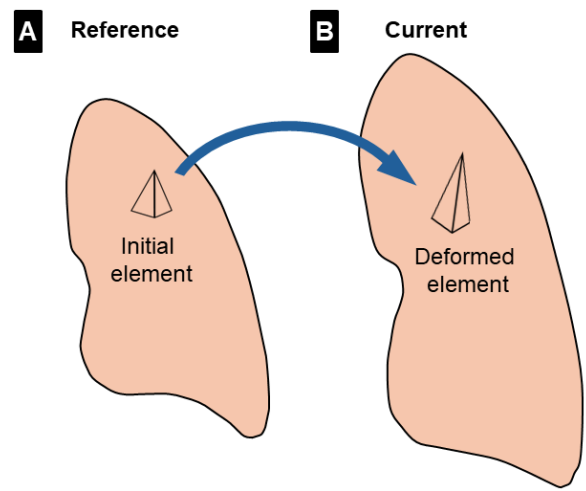

Fig. 1. Schematic of the (A) initial geometry of the lung and a representative element within the geometry. (B) The deformed geometry of the lung after inspiration, and the deformed and translated element

Lung image registration is used to estimate the displacement and deformation experienced by the lungs during respiration [8]-[13]. These kinematics metrics can be used to estimate mechanistic markers of regional deformation, such as volumetric expansion and distortion, which can potentially act as surrogates of ventilation and shearing in the alveolar septal walls, respectively. However, a limiting factor in the clinical translation of image registration-based mechanistic markers is the need for images with a sufficiently high temporal resolution corresponding to small displacements between successive timepoints. Dynamic CT scans can acquire images at the required spatiotemporal resolution and have been used to investigate the kinematics of parenchymal tissue in animal models [7], [9]. However, dynamic CT scans are limited in clinical applications due to the excessively high dosages of radiation associated with dynamic imaging.

This work was supported by the National Institutes of Health (R00HL138288 to R.A.)

Inspiratory static CT scans acquired at end-inspiration (EI) are commonly used in conjunction with expiratory CT scans acquired at end-expiration (EE) for diagnostic purposes in clinics. Since these are static scans, they can be performed at high spatial resolution, providing significant morphological detail in the images. However, the large displacements between these image stacks limit the effectiveness of image registration, resulting in a need for a method that can predict the deformation of lung parenchyma between two timepoints (i.e., EI and EE) that are spatially distant. In-silico modeling has emerged as a powerful tool for investigating biomechanical behavior in soft tissues such as the lungs [14]. In-silico modeling, used in conjunction with medical imaging, can predict kinematic behavior in the lungs based on the available, yet sizably distinct, geometries at EE and EI. In this study, we developed a rat-specific finite element (FE) method with physiologically realistic loading to estimate lung deformation using only image stacks at the EE and EI timepoints (Fig. 1). We evaluated the FE prediction against the image registration results using the high-temporal resolution dynamic images in the same rat.

## II. METHODOLOGY

*A. Dynamic imaging and image registration*

We previously used an image registration algorithm for kinematic quantification in rat lungs using dynamic CT [15]. Briefly, four-dimensional (4D) CT scans were obtained for a healthy Sprague Dawley (SD) rat placed under isoflurane anesthesia. Sixteen equally spaced image stacks were collected over the course of one respiratory cycle and used for image registration. We used an open-source image registration code (Nifty-reg) to perform non-linear, non-rigid image registration to obtain voxel-based displacement during respiration and, subsequently, the deformation. The lungs were segmented from the reference image stack using 3D Slicer, and the segmentation was propagated through image registration to track the region of interest throughout the inspiration cycle. From the displacement $\mathbf{u} = [u_x, u_y, u_z]$, the linear-elasticity strain ($\varepsilon$) was estimated using the following equations

$$\mathbf{F} = \mathbf{I} + \begin{bmatrix} \frac{\partial u_x}{\partial x} & \frac{\partial u_y}{\partial x} & \frac{\partial u_z}{\partial x} \\ \frac{\partial u_x}{\partial y} & \frac{\partial u_y}{\partial y} & \frac{\partial u_z}{\partial y} \\ \frac{\partial u_x}{\partial z} & \frac{\partial u_y}{\partial z} & \frac{\partial u_z}{\partial z} \end{bmatrix}, \quad (1)$$

$$\varepsilon = \frac{1}{2}(\mathbf{F^T} + \mathbf{F}) - \mathbf{I}, \quad (2)$$

where $\mathbf{I}$ is the identity tensor.

*B. Image reconstruction at EE and EI timepoints*

Individual images acquired at the EE and EI timepoints from dynamic imaging were used to segment and reconstruct the lung geometry as the reference and current geometries, respectively (Fig. 2). The volume of the reference (EE) and current (EI) geometries were 5.4 cm³ and 7.8 cm³, respectively. The images were segmented in 3D Slicer, and the geometries were reconstructed using Materialise 3-matic. The reconstructed geometries were meshed using linear tetrahedral elements to create a volumetric description of the lung at both timepoints. The element sizes were kept constant between the two volumetric meshes, but the number of elements and nodes were different.

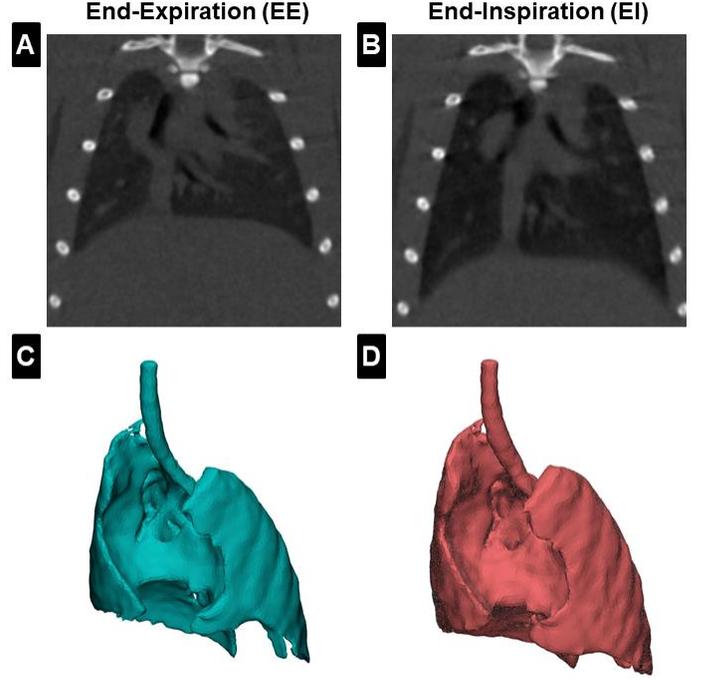

Fig. 2. Dynamic computed tomography (4D-CT) image from rat at (A) end-expiration (EE) and (B) end-inspiration (EI). Corresponding lung geometries segmented and reconstructed at the (C) EE and (D) EI timepoints

*C. Finite element simulation*

The lungs were modeled as a compressible linear elastic material in the FE model. The reference mesh was fixed at the trachea and the region where the primary bronchi attach to the lungs. Two normal force distributions were applied, one to the surface of the lungs in contact with the diaphragm and the second on the surface of the lungs in contact with the ribcage. The magnitude of the forces was estimated by minimizing the difference between the reference (EE) and current (EI) mesh. Following the first step of the simulation, an additional second step was performed where displacement boundary conditions were applied to further minimize the difference between the surfaces of the two geometries. The steps with the force and displacement boundary conditions serve as a coarse and fine step in minimizing the error between the reference and current meshes.

*D. Error estimation between two FE meshes*

The difference between the predicted and ground truth EI geometry was quantified by the Hausdorff distance ($d_H$) [16], which is given by

$$d_H(X,Y) = max\left\{\sup_{x \in X}\inf_{y \in Y} d(x,y), \sup_{y \in Y}\inf_{x \in X} d(x,y)\right\}, \quad (3)$$

where sup and inf are the supremum and infimum functions, respectively. The supremum of a subset is defined as the smallest element that is greater than all the elements in the subset, while the infimum is the greatest element that is smaller than all the elements of the subset. $x \in X$ and $y \in Y$ are nodes belonging to mesh $X$ and $Y$, respectively. To calculate the Hausdorff distance, the distance between one node of mesh $X$ and all nodes of mesh $Y$ were calculated, and the minimum distance was noted. This process was repeated for all nodes of mesh $X$, and the maximum distance was recorded. Next, the process was iterated by estimating the minimum distance from each node of mesh $Y$ to the nodes of mesh $X$, and finding the maximum such distance for all nodes in mesh $Y$. The overall maximum distance was taken as the Hausdorff distance between the two meshes. Once the Hausdorff distance was minimized through the FE simulation, the strain was estimated from the EI mesh estimated by the FE method. The final Hausdorff distance was normalized against the characteristic height of the lungs at EE (40mm.)

## III. RESULTS

The results indicated that the geometry obtained through the in-silico simulation matched closely with the geometry segmented and reconstructed from the image acquired at EI (Fig. 3). Visualizing the Hausdorff distance indicates that the error at the "edges", where the curvature of the geometry changes sign, is high (Fig. 3A,B). The Hausdorff distance was 0.6 mm at the inner surface of the lungs, and 0.25mm at the outer surface of the lungs. The error metric, defined as the ratio of Hausdorff distance to the height of the lungs (40 mm for the reference geometry), was 2.275% for the deformed geometry.

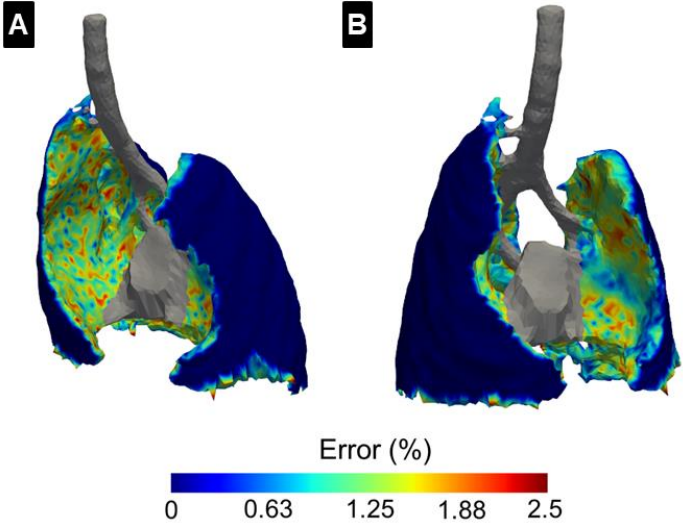

Fig. 3. Contour map of the error defined as the ratio of Hausdorff distance to characteristic z-length (40mm) across the surface of the lungs in two views. The contour was mapped onto the EI mesh estimated by the FE method.

The simulation-derived strain along the z-direction, averaged over the entire lungs, was $\varepsilon_{zz}$=0.08±0.24. In comparison, the corresponding strain from image registration was found to be $\varepsilon_{zz}$=0.12±0.28. Better agreements were found for regional strains. For instance, when averaging was restricted to a transverse plane (Fig. 4A-inset), the strains from the FE method and image registration were $\varepsilon_{zz}$=0.111±0.196 (Fig. 4A) and $\varepsilon_{zz}$=0.118±0.128 (Fig. 4B), respectively. For the transverse section, the lateral strains estimated by the FE method were $\varepsilon_{xx} = -0.02 \pm 0.13$ and $\varepsilon_{yy} = -0.007 \pm 0.12$, as compared to $\varepsilon_{xx} = 0.05 \pm 0.21$ and $\varepsilon_{yy} = 0.008 \pm 0.15$ estimated by image registration. These results indicate that the FE method can match the results estimated by image registration along the primary displacement (z) direction. The difference in the global mean results estimated by the FE and image registration is primarily linked to the limitation of the proposed method in accurately estimating and imposing boundary conditions along the edge of the surface of the lungs in contact with the diaphragm. The applied displacement boundary conditions will be refined in future studies. The displacement boundary condition on the diaphragm will be refined to ensure that the diaphragm remains a smooth surface to more accurately replicate physiological respiration.

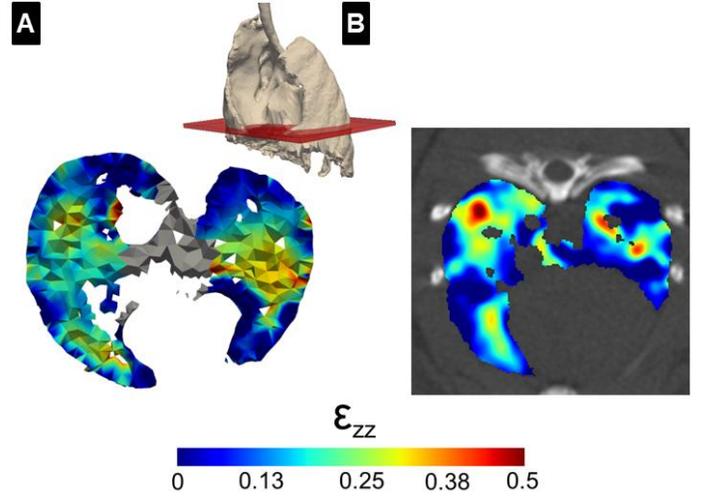

Fig. 4. $\varepsilon_{zz}$ of an isolated section at end-inspiration relative to end-expiration estimated by (A) the FE methods and (B) image registration. The inset in (A) highlights the location of the section. The region near the sternum was greyed out to exclude artifacts caused by the boundary conditions.

## IV. DISCUSSION

In this study, we have presented an FE method to recapitulate physiologically realistic loading in the lungs to estimate large displacements between two statically acquired CT image imaging techniques. We estimated node-wise displacement and element-wise deformation through a combination of segmentation, reconstruction, and FE modeling. The Hausdorff distance indicated a minimal difference between the deformed mesh obtained through reconstruction and FE simulation. For context, when the reference mesh is scaled by 2% (multiplied by 1.02), the maximum Hausdorff distance is 0.91 mm (2.275% when divided by 40mm, the characteristic length along z). These results indicate that the proposed physics-informed FE method can predict the displacements mapping lung morphology from EE to EI. The strain measures indicated that the FE method can capture the dominant deformation (along the vertical direction) but has lower accuracy in the lateral directions (x-direction). The change of surface concavity along the edge of the surface of the lungs in contact with the diaphragm leads to errors when calculating the Hausdorff distance. The estimation of the Hausdorff distance and application of boundary conditions need to be refined to minimize the relatively large errors observed at these edges of the lung geometry. In addition, the prediction image registration needs to be validated in and of itself. We have developed an in-silico-based

CT image phantom of the lungs [17] that can be set up to evaluate the image registration technique predictions in future studies.

In addition to the ability to work with static images, a key advantage of the method presented in this study over conventional image registration is the ability to determine the exact deformation of any given tetrahedral element. In essence, the FE method can estimate the displacement, rotation, and deformation of a region of interest in the volume of the lung parenchyma, which can prove to be essential in the case of heterogeneous lung injuries such as inflammation, VILI, and RILI. Since such lung diseases also cause regional variations in parenchymal elasticity, our proposed method can be extended to ensure the mapping of not only the lung borderlines but also any interior region that may have a different elasticity than that of healthy tissues. Another key advantage of such a method over image registration is the capability to use images with lower spatial resolution. Since the method estimates displacement through FE simulation, strains can be accurately estimated by matching the displacement of the lung surface at key landmarks such as large airways. This capability of working with lower-resolution images and estimating mechanistic biomarkers from static CT images can significantly enhance the potential application of biomechanical markers in clinics as a diagnostic tool for risk stratification and individualized treatments.

Despite several benefits of the presented method, it has some limitations at its current stage, including a high computational cost. In particular, if fine details of the lung structure, such as distal airways, are segmented and reconstructed, the number of nodes and elements of the mesh might become excessively high. A large number of nodes and elements would lead to computationally expensive in-silico simulations, thus significantly increasing the time required to estimate displacement. However, one potential solution would be to leverage the method through machine learning (ML) techniques. While computationally intensive during training, ML has proved invaluable in reducing computational expenses during run time. While studies have used deep learning and neural networks to perform image registration of the lungs [9], [13], deep learning has also been used to estimate the results from in-silico simulations based on parameters such as geometry, material behavior, and loading conditions [18]. Future studies and extensions will involve using machine learning to reduce computational costs during run time to enhance the translational potential of clinical applications. Additionally, the FE model used in this study was limited to linear elastic material behavior. Future studies will implement hyperelastic material behavior and Green-Largrange strain definition instead of $\epsilon$ for more accurate estimations of the strains.

## V. CONCLUSIONS

The method presented in this study offers the possibility of estimating mechanical markers, such as stress and strains, using static medical imaging acquired at EE and EI timepoints. We expect this method to serve as an essential step in promoting the feasibility of using regional kinematic biomarkers, such as strain, in diagnostic and prognostic assessment of lung injury patients.

## ACKNOWLEDGEMENT

This work was supported by the National Institutes of Health R00HL138288 to R.A.